\documentclass[conference]{src/IEEEtran}
\IEEEoverridecommandlockouts
\usepackage{cite}
\usepackage{amsmath,amssymb,amsfonts}
\usepackage{algorithmic}
\usepackage{hyperref}
\usepackage{graphicx}
\usepackage{textcomp}
\usepackage[table]{xcolor}
\usepackage{xcolor}
\usepackage{microtype}
\usepackage{booktabs,multirow}
\usepackage{tabularx} 
\usepackage{makecell} 
\usepackage{multirow} 
\usepackage{booktabs} 
\usepackage{multicol}

\newcommand{\argmin}{\operatornamewithlimits{argmin}}

\def\BibTeX{{\rm B\kern-.05em{\sc i\kern-.025em b}\kern-.08em
    T\kern-.1667em\lower.7ex\hbox{E}\kern-.125emX}}
    
\begin{document}




\title{Mind the Domain Gap: a Systematic Analysis on Bioacoustic Sound Event Detection}
\author{
\IEEEauthorblockN{
Jinhua Liang\textsuperscript{$\dagger$} \quad 
Ines Nolasco\textsuperscript{$\dagger$} \quad 
Burooj Ghani\textsuperscript{$\diamond$} \quad
Huy Phan\textsuperscript{$\ddagger$}\thanks{\textsuperscript{$\ddagger$}The work does not relate to Huy Phan's work at Amazon.}\quad 
Emmanouil Benetos\textsuperscript{$\dagger$} \quad 
Dan Stowell\textsuperscript{$\ast$}\textsuperscript{$\diamond$}
}
\\
\IEEEauthorblockA{\textsuperscript{$\dagger$}\textit{Centre for Digital Music, Queen Mary University of London, London, UK}\\
\textsuperscript{$\diamond$}\textit{Naturalis Biodiversity Center, Leiden, Netherlands}\\
\textsuperscript{$\ast$}\textit{Department of Cognitive Science and Artificial Intelligence, Tilburg University, Tilburg, Netherlands}\\
\textsuperscript{$\ddagger$}\textit{Amazon AGI, Cambridge, UK}}

}


\maketitle

\begin{abstract}
Detecting the presence of animal vocalisations in nature is essential to study animal populations and their behaviors. A recent development in the field is the introduction of the task known as few-shot bioacoustic sound event detection, which aims to train a versatile animal sound detector using only a small set of audio samples. Previous efforts in this area have utilized different architectures and data augmentation techniques to enhance model performance. However, these approaches have not fully bridged the domain gap between source and target distributions, limiting their applicability in real-world scenarios. In this work, we introduce an new dataset designed to augment the diversity and breadth of classes available for few-shot bioacoustic event detection, building on the foundations of our previous datasets. To establish a robust baseline system tailored for the DCASE 2024 Task 5 challenge, we delve into an array of acoustic features and adopt negative hard sampling as our primary domain adaptation strategy. This approach, chosen in alignment with the challenge's guidelines that necessitate the independent treatment of each audio file, sidesteps the use of transductive learning to ensure compliance while aiming to enhance the system's adaptability to domain shifts. Our experiments show that the proposed baseline system achieves a better performance compared with the vanilla prototypical network. The findings also confirm the effectiveness of each domain adaptation method by ablating different components within the networks. This highlights the potential to improve few-shot bioacoustic sound event detection by further reducing the impact of domain shift.
\end{abstract}


\section{Introduction} \label{sec:introduction}


The automatic detection and analysis of bioacoustic sound events is a valuable tool for assessing animal populations and their behaviours, facilitating conservation efforts, and providing insights into ecological dynamics~\cite{caiger2020decade, gillings2021nocturnal, tuia2022perspectives}. By leveraging advanced machine learning and computational bioacoustics techniques, it is possible to decode the complex acoustic signals of diverse species~\cite{ganchev2017computational}. This approach can allow us to overcome some of the challenges to a considerable extent, such as background noise, simultaneous calls from various sources, and the major issue of limited annotated data in the face of an expanding volume of recordings~\cite{stowell2022computational}. Despite the advancements, the domain-specific challenges of bioacoustic sound event detection – such as distinguishing between various vocalisation types, dialect types, species versus individuals identification, models capable of decoding both soundscapes and focal recordings – underscore the need for general-purpose systems that are capable of adapting to these specialized tasks~\cite{ghani2023global, hamer2023birb, boudiaf2023search}. Addressing these fragmentation and specificity challenges in bioacoustic sound event detection, few-shot learning (FSL) emerges as a promising framework capable of transcending the varied subdomains within computational bioacoustics~\cite{nolasco2023learning,Li2022-3}. FSL in sound event detection leverages the concept that a system can be trained on a small, representative set of data samples to detect and classify novel, unseen sound events. By training a single model to adapt to multiple datasets with minimal examples, FSL proposes a robust and versatile solution~\cite{Liang2022,liang23c_interspeech}. This innovation not only addresses the lack of large-scale annotated datasets but also the practicality of applying deep learning techniques across the fragmented landscape of bioacoustic research, offering a pathway to more scalable and efficient detection systems~\cite{nolasco2023learning}. 


To diversify methods and accelerate progress in FSL within the bioacoustics domain, an open-science task was organized within the Detection and Classification of Acoustic Scenes and Events (DCASE) framework, focusing on bioacoustic sound event detection (SED). This challenge, open to all participants, featured public datasets, evaluation metrics, documentation, and baseline systems~\cite{morfi2021few,nolasco2022few,nolasco2023few}. The few-shot bioacoustic sound event detection task highlights some unique features: 1) diverse target sounds: sounds of different animals vary greatly due to the distinct mechanism of their vocalisation; 2) sparse distribution of animal vocalisations: the presence of animal sounds is much rarer than the background slices in the recording; and 3) dynamic acoustic environment: the context of target sounds changes drastically due to the heterogeneous nature of various microphones, different sound sources, and environments. We assume these unique features pose a challenge of domain shift, hindering the generalization ability of few-shot bioacoustic detectors.



In this work we explore the problem of domain shift in the field of bioacoustics and present a new baseline system for the few-shot bioacoustic event detection task in the DCASE 2024 challenge\footnote{\url{https://dcase.community/challenge2024/}}. This is achieved by adapting the system proposed in~\cite{Liu2022a}. Diverse acoustic features are investigated and two domain adaptation techniques, namely, negative hard sampling~\cite{robinson2020contrastive} and transductive learning~\cite{arnold2007comparative} are systematically ablated to study their impact on the performance on the few-shot bioacoustic event detection task. Additionally, to mitigate the domain shift, we introduce a new dataset for DCASE Task 5 for 2024 by extending the dataset from the previous years challenges. We conduct an additional ablation study and compare the performance on different versions of the dataset to highlight the challenge of domain shift and analyse the optimum machanisms for a good performance in this task.
In adherence to the DCASE 2024 Task 5 challenge rules, which require treating each audio file independently to prevent potential class overlap in the evaluation set, our baseline system does not incorporate transductive learning - a technique we evaluated in our 2022 analysis. Instead, our domain adaptation efforts for the 2024 challenge are centered around negative hard sampling, ensuring compliance with the challenge's guidelines and focusing on methodologies conducive to advancing few-shot bioacoustic event detection research within the stipulated framework.



\section{Dataset} \label{sec:dataset}

\begin{table*}[h]
\centering
\caption{Summary of dataset characteristics.}
\label{tab:datasets}
\resizebox{\textwidth}{!}{%
\begin{tabular}{@{}r|l|c|c|c|c|c|c@{}}
\toprule
                    & Name and species                           & Mic type       & \# Audio files & Total duration & \# Labels & \# Events & Mean Event duration (s) \\ \midrule
                    & BV: BirdVox-DCASE-10h                      & fixed          & 5              & 10 hours       & 11        & 9026      & 0.15                       \\
                    & HT: Hyenas                                 & various        & 5              & 5 hours        & 5         & 611       & 1.42                       \\
Training set        & MT: Meerkats                               & animal mounted & 2              & 70 mins        & 4         & 1294      & 0.14                       \\
                    & JD: Jackdaws                               & mobile         & 1              & 10 mins        & 1         & 357       & 0.12                       \\
                    & WMW: Western Mediterranean Birds           & various        & 161            & 5 hours        & 26        & 2941      & 1.54                       \\ \midrule
                    & HB: Humbug mosquitoes                      & handheld       & 10             & 2.38 hours     & 1         & 712       & 11.6                       \\
Validation set 2022 & PB: Polish Baltic Sea bird flight calls    & fixed          & 6              & 3 hours        & 2         & 292       & 0.11                       \\
                    & ME: Meerkats                               & animal mounted & 2              & 20 mins        & 2         & 73        & 0.19                       \\ \midrule
                    & Validation set 2022 +                      &                &                &                &           &           &                            \\
                    & PB24:  Polish Baltic Sea bird flight calls & fixed          & 4              & 120 mins       & 2         & 350       & 0.08                       \\
Validation set 2024 & RD: Red Deer                               & fixed          & 6              & 18 hours       & 1         & 1372      & 1.52                       \\
                    & PW: Pilot Whales                           & animal mounted & 15             & 24 hours       & 1         & 705       & 2.21                       \\ \bottomrule
\end{tabular}%
}
\vspace{-0.5cm}
\end{table*}

In this work we make use of the development dataset from the DCASE 2022's few-shot bioacoustic event detection task~\cite{nolasco2022few}.
As described in \cite{nolasco2023learning}, each subset of data represents different bioacoustic sources and contains long duration recordings of different species, recorded in various locations and with different recording setups and devices. Due to this diversity, the data is able to represent the nature of the bioacoustics domain which is commonly described more as a set of computational related tasks instead of single one.
The training set recordings are paired with multi-class annotations while in the validation set each audio recording is only annotated for a single class of interest. This means that in the validation set, other salient/foreground events exist however these are not the target of the detection task. 
To further illustrate the heterogeneous nature of the data, we note that the level of the annotations can also vary across the different subsets. In some, the target classes are different species (see table \ref{tab:datasets} for WMW, BV and PB), however in other sets the target classes are different call types of the same species (see table \ref{tab:datasets} for ME).

A summary description of the different characteristics in the dataset is presented in Table \ref{tab:datasets}.

Focusing on the domain shift problem, this dataset provides a realistic example of the expected degree to which data distributions can shift and vary across the different sets of data.
In fact the training set has minimum overlap between classes and characteristics of the data in the validation set which constitutes a challenging scenario for system development, particularly since the expectation of similar performance on training and validation sets cannot be guaranteed here. 
To further highlight the occurrence of domain shifts in bioacoustic datasets, we also consider the recently updated version of the validation set\footnote{\url{https://doi.org/10.5281/zenodo.10829604}}, see Table \ref{tab:datasets}, Validation set 2024.
The validation set has been extended to include more flight calls recordings (PB data) and recordings of two new species: Red Deer (RD) and Pilot Whales (PW).
It is also of note that with the addition of Pilot whale recordings in the validation set, we are introducing an important domain shift since the acoustic underwater soundscape is completely different from the other acoustic environments.


\section{Proposed system} \label{sec:proposed_system}
To automatically detect the presence of an animal in a recording, the bioascoutic sound event detection system should learn the concept of the vocalisation of a specific species from a limited number of references. This poses a unique challenge in the task of few-shot bioacoustic sound event detection: the target domain could be misaligned with the source domain from which the model learns the concept of animal vocalisation. In Sect.~\ref{subsec:problem_formulation} we formulate the problem of few-shot bioacoustic sound event detection and introduce the architecture of the proposed baseline in Sect.~\ref{subsec:model_architecture}. To mitigate the domain gap between a source and a target domain, we elaborate two designs in Sect.~\ref{subsec:domain_adaptation}. 

\subsection{Problem formulation} \label{subsec:problem_formulation}
Suppose a dataset $\mathcal{D} = \{(\mathbf{x}_i, \mathbf{y}_i)\}_{i=1}^{|\mathcal{D}|}$ where $\mathbf{x}_i$ denotes the $i$-th audio example, $\mathbf{y}_i \subset \mathcal{C}$ denotes the set of discrete labels of the $i$-th example, and $\mathcal{C}$ denotes the label set of $|\mathcal{C}|$ classes, $\mathcal{C}=\{1,\ldots,|\mathcal{C}|\}$. 

For two non-overlapping subsets $\mathcal{D}_s,\mathcal{D}_q\subset \mathcal{D}$, a classical few-shot learning setting can be formulated as a ``$N$-way $K$-shot'' problem where a classification task is constituted by: (i) a label subset $\mathcal{C}_s$ of $N$ classes sampled from $\mathcal{C}$, (ii) $K$ examples (known as \textit{support} data) sampled from $\mathcal{D}_s$ for each class in $\mathcal{C}_s$, and (iii) $Q$ examples (known as \textit{query} data) sampled from $\mathcal{D}_q$ for each class in $\mathcal{C}_s$. With a model family chosen as $p_\theta(\mathbf{x})$, with unknown parameters $\theta\in\Theta$, the problem boils down to maximizing the average likelihood of all the samples from $\mathcal{D}_q$ under the model parameter:
\begin{equation} \label{eqn:problem_formulation}
    \theta^*=\argmin_{\theta\in\Theta}\frac{1}{|\mathcal{D}_q|}\sum\limits_{(\mathbf{x}, \mathbf{y})\in\mathcal{D}_q}\mathcal{L}(\mathbf{y}|\mathbf{x},\theta)
\end{equation}

\subsection{Model architecture} \label{subsec:model_architecture}
We adopt prototypical networks~\cite{snell2017prototypical} to directly learn the concept of animal vocalisation in the latent space. For a class $n\in\mathcal{C}_s$, let $\mathcal{S}_{n}$ be the subset of support examples containing the sound events of this class, $|\mathcal{S}_{n}|= K$. Then the set of sound events is represented as $\mathcal{S}_{n}^+\subset\mathcal{S}_{n}$ where the set of negative segmentation is denoted as $\mathcal{S}_{n}^-=\mathcal{S}_{n}/\mathcal{S}_{n}^+$.

The positive prototype $\mathbf{a}_n^+$ of class $n$ is then derived as the mean of embedding vectors of the support examples in $\mathcal{S}_{n}^+$.
Formally, 
\begin{equation}
    \label{eq:protonet_pos_prototype}
    \mathbf{a}_n^+=\frac{1}{K}\sum\limits_{(\mathbf{x},\mathbf{y})\in \mathcal{S}_{n}^+} f_{\theta}(\mathbf{x}),
\end{equation}
where $f$ denotes the embedding mapping realized by the model whose parameters are denoted collectively as $\theta$. Likewise, the negative prototype $\mathbf{a}_n^-$ of class $n$ is calculated by
\begin{equation}
    \label{eq:protonet_neg_prototype}
    \mathbf{a}_n^-=\frac{1}{K}\sum\limits_{(\mathbf{x},\mathbf{y})\in \mathcal{S}_{n}^-} f_{\theta}(\mathbf{x}),
\end{equation}
Given a query example $\mathbf{x}_q$, the model performs $N$ binary classifications independently by producing a probability distribution based on a softmax over distances between $\mathbf{x}_q$ and the positive and negative prototypes in the latent space. More specifically, the probability that $\mathbf{x}_q$ is classified as class $n\in\mathcal{C}_s$ is calculated as 
\begin{equation}
    \label{eq:protonet_loss} 
    p_{\theta}(\hat{\mathbf{y}}_{q,n} | \mathbf{x}_q)=\frac{e^{-d(f_{\theta}(\mathbf{x}_q), \mathbf{a}_n^+)}}
    {\sum\limits_{i\in\mathcal{C}_s} \left[e^{-d(f_{\theta}(\mathbf{x}_q), \mathbf{a}_i^+)}+e^{-d(f_{\theta}(\mathbf{x}_q), \mathbf{a}_i^-)}\right]},
\end{equation}
where $\hat{\mathbf{y}}_{q,n}$ is the predicted label for $\mathbf{x}_q$ w.r.t. class $n$, $d$ is a distance function, such as $\ell_2$ or cosine distance. The network is trained to minimize the negative log-probability of the true class over the $N\!\times\!Q$ query examples: 

\begin{equation}
    \label{eq: protonet_celoss}
    \mathcal{L}(\theta) = \sum\limits_{(\mathbf{x},\mathbf{y})\in \mathcal{Q}}\sum\limits_{n\in\mathcal{C}_s}-\log p_{\theta}(\hat{\mathbf{y}}_{n}=\mathbf{y}_{n} | \mathbf{x}),
\end{equation}
where $\mathcal{Q}$ is the set of query examples, $|\mathcal{Q}|\!=\!N\times Q$.

While prototypical networks perform well in few-shot learning settings~\cite{Liang2024}, their performance still suffers when the domain of the development dataset is drastically different from the one of the test dataset. To mitigate the domain misalignment issue, we adopt two strategies in the following part.

\subsection{Domain adaptation} \label{subsec:domain_adaptation}
\textbf{Negative hard sampling}.
As shown in eqn.~(\ref{eq:protonet_loss}), both positive and negative prototypes are important to predict the probability distribution over each class. Bioacoustic sound events are sparsely distributed throughout a recording. While these sound events are hardly available during training, there are many negative audio segments available. This drives us to investigate how to create ``meaningful'' negative prototypes by using hard sampling techniques~\cite{robinson2020contrastive} during inference stage. Instead of using all negative audio segments, we randomly sample a proportion of segmentations for negative prototype measurement. In other words, we inject the randomness to the cluster of negative segments such that the trained model can generalise to evaluation set well. In addition, we enhanced the trained prototypical network with different set of negative segments to further improve the performance of the proposed model on the evaluation set.

\textbf{Transductive learning}.
Acoustic environment of test recordings may vary from the development domain due to many factors, such as the difference of recording contexts, recording devices, and pre-processing methods. To alleviate the domain mismatch caused by different acoustic environments, we apply transductive inference~\cite{arnold2007comparative} by learning not only from the labelled events of the training set but the first five labelled examples for each audio recording in the validation set. 

\begin{table*}[]
\centering
\caption{Benchmarking the baseline system (\%) on DCASE 2022 Task 5 and DCASE 2024 Task 5 validation sets. The gray bar highlights the performance score of our proposed baseline for DCASE 2024 Task 5.}
\label{tab:main_result}
\begin{tabular}{@{}ccccc|ccc@{}}
\toprule
\multirow{2}{*}{Negative hard sampling} & \multirow{2}{*}{Transductive learning} & \multicolumn{3}{c|}{DCASE 2022 Task 5}                           & \multicolumn{3}{c}{DCASE 2024 Task 5} \\
                                        &                                        & Precision          & Recall             & $F_1$-score        & Precision   & Recall  & $F_1$-score  \\ \midrule
                                        &                                        & 55.92±2.6          & 40.61±5.6          & 46.78±2.9         & 44.94±3.39         & 45.89±4.80         & 45.23±0.48         \\
                                        &   \checkmark                               & 55.17±3.6          & 43.59±0.4          & 48.66±1.6          &             &         &           \\
                        \rowcolor{gray!30} \checkmark  &                                        & 48.33±2.3          & 56.64±2.0          & 52.09±0.7          & \textbf{56.18±0.61}            & \textbf{48.64±0.23}        & \textbf{52.14±0.20}          \\
                            \checkmark    &  \checkmark                            & \textbf{66.43±3.6} & \textbf{61.28±1.5} & \textbf{63.67±1.0} &             &         &           \\ \bottomrule
\end{tabular}
\vspace{-0.5cm}
\end{table*}

\begin{table}[]
\centering
\caption{Performance comparison (\%) of systems with diverse acoustic features on the DCASE 2022 Task 5 validation set.}
\label{tab:compare_features}
\begin{tabular}{@{}lccc@{}}
\toprule
Feature              & Precision          & Recall             & $F_1$-score        \\ \midrule
Mel                  & 38.56±1.2          & 49.01±1.5          & 43.14±0.9          \\
Log mel              & 66.43±3.6          & 61.28±1.5          & \textbf{63.67±1.0} \\
Log mel + MFCC       & 58.89±1.1          & \textbf{65.70±1.0} & 62.10±0.5          \\
Log mel + delta MFCC & 61.22±1.4          & 62.75±0.6          & 61.96±0.7          \\
PCEN                 & \textbf{68.00±2.2} & 53.70±1.7          & 59.97±0.7          \\
PCEN + MFCC          & 63.88±1.7          & 57.65±0.7          & 60.59±0.6          \\
PCEN + delta MFCC    & 59.47±5.1          & 52.22±1.8          & 55.47±1.5          \\ \bottomrule
\end{tabular}
\vspace{-0.5cm}
\end{table}

\section{Experiment} \label{sec:experiment}
\subsection{Experiment setup} \label{subsec:experiment_setup}
For a fair comparison, we adopted the same convolutional neural network as the backbone for feature extraction. Following~\cite{Liu2022a}, this network consists of three 3$\times$3 convolutional layers, with dimension of 64, 128, 64. Each of the convolutional layer is followed by batch normalisation, leak ReLU activation, and a 2$\times$2 max-pooling layer. For each convolutional block, we added a skip connection followed by a max-pooling layer to avoid catastrophic forgetting.

We trained and evaluated the models on DCASE 2022 Task5 and DCASE 2022 Task5 datasets, respectively. For all experiments, we resampled audio recordings to 22.5 kHz sampling rate and applied short-time Fourier transformation (STFT) with 1024 of window length and 256 of hop length. We set the number of frequency bins as 128 to extract log-Mel spectrograms (Log Mel) from audio examples. To investigate the impact of acoustic features on this task, we also experimented diverse acoustic features, including the Mel spectrogram (Mel), Mel-frequency cepstral coefficients (MFCC), delta features of MFCC (delta MFCC), per-channel energy normalization (PCEN), and their combinations as input features. Specifically, the dimension of MFCC features is set to 32 in the experiments.

We trained the models using RTX A5000 GPUs. Following~\cite{Liu2022a}, we set the initiate learning rate to 0.001 with exponential decay of 0.65 for every 10 epochs. We adopted early stopping by monitoring the accuracy score and took the checkpoint of the highest accuracy score as our best model. During evaluation, we fixed the set of negative audio segments across models with the same number of negative segments. For each experiment, we calculated the average performance scores with 95\% confidence score over five independent trials to avoid impact of randomness. Our code and implementation can be found at this url\footnote{\url{https://github.com/c4dm/dcase-few-shot-bioacoustic/tree/main/baselines/dcase2024\_task5}}.

\subsection{Experimental results} \label{subsec:experiment_results}
Table~\ref{tab:main_result} details the baseline system's performance on the DCASE 2022 Task 5 and DCASE 2024 Task 5 validation sets, alongside an ablation study for domain adaptation techniques applied to the DCASE 2022 Task 5 dataset. The baseline system achieved a precision of 48.33\%, a recall of 56.64\%, and an F1-score of 52.09\% for DCASE 2022 Task 5, and a precision of 56.18\%, a recall of 48.64\%, and an F1-score of 52.14\% for DCASE 2024 Task 5. We explored integrating transductive learning with the baseline system, which showed improved performance with an increase in precision by 18.1\%, recall by 4.64\%, and F1-score by 11.58\% absolute, indicating the impact of domain shift in few-shot bioacoustic sound event detection.

Given the DCASE task rules, which stipulate treating each audio file independently so that each item in the evaluation set is handled as an independent few-shot scenario, we adhered to these guidelines in our approach to the DCASE 2024 Task 5 challenge. Consequently, the version of transductive learning we used for our 2022 analysis was not applied for the 2024 dataset. For the DCASE 2024 Task 5, we focused our domain adaptation efforts on negative hard sampling, excluding transductive learning from our methodology. This strategic decision reflects our commitment to adhering to the DCASE task rules while still advancing research in domain adaptation for few-shot bioacoustic sound event detection; various forms of domain adaptation remain as future research possibilities.


Table~\ref{tab:compare_features} compares diverse acoustic features on DCASE 2022 Task 5 validation set. The system with log Mel features achieves the best $F_1$-score of compared to those of other acoustic features. Among these models, the combination of log Mel and MFCC yielded the best recall score of 65.70\% while PCEN had the best precision score of 68.00\%. Provided that we expect a balanced performance of bioacoustic sound event detectors, we applied log Mel as the input acoustic features for the baseline. In addition, the precision scores of log Mel and PCEN are lower than their counterpart combined with MFCC, respectively. This probably indicates that equipping input features with MFCC is beneficial to models' performance in terms of the precision score.

\begin{figure}
    \centering
    \includegraphics[width=0.9\columnwidth]{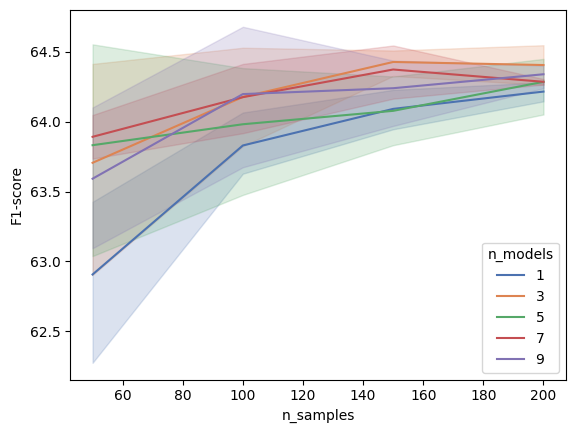}
    \vspace{-0.4cm}
    \caption{Ablation study on the number of negative audio segments and number of enhanced models.}
    \label{fig:neg_plot}
    \vspace{-0.5cm}
\end{figure}

Fig.~\ref{fig:neg_plot} illustrates the ablation study on the number of negative audio segments and enhanced models. Enhanced models hereby are referred to as the model of the same checkpoint with different set of negative segments. As the number of sets of negative audio segments increases, the system' performance varies less. Likewise, as the number of negative segments increases, the performance score of systems becomes more stable. In addition, the system of three models with 150 negative segments achieved the best $F_1$-score performance score. This guides us to create the baseline system of the DCASE 2024 Task 5 dataset.

\begin{figure}
    \centering
    \includegraphics[width=1\columnwidth]{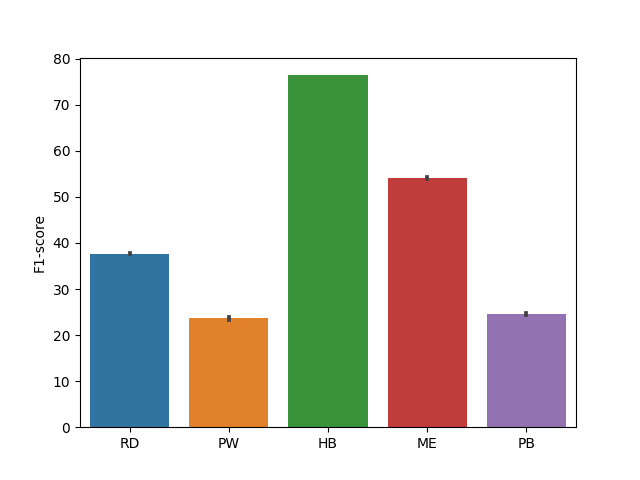}
    \vspace{-1cm}
    \caption{Species-wise performance of the baseline on DCASE 2024 Task 5 validation set.}
    \label{fig:species-wise_performance}
    \vspace{-0.5cm}
\end{figure}

Fig.~\ref{fig:species-wise_performance} compares the performance of the baseline on DCASE 2024 Task 5 validation set by species. It can be observed that the baseline yielded the best performance on the species of humbug mosquitoes while performing the worst on that of pilot whales. The latter observation is in line with our expectation: the model did not learn any underwater sounds from the training set and thus cannot be adapted to the vocalisation of marine animals well.

\section{Conclusion} \label{sec:conclusion}
In this work, we benchmarked the task of few-shot bioacoustic sound event detection using the recently updated DCASE 2024 Task 5 dataset. We pinpointed domain shift as a significant challenge within this task. To substantiate our hypothesis, we improved on the previous dataset, aimed at enhancing the diversity and enlarging the size of the preceding version. We experimented with a variety of acoustic features as inputs for the baseline model to assess the effects of feature engineering on bioacoustic computations. Additionally, we improved the prototypical network by integrating negative hard sampling and transductive learning, aiming to better adapt the network to the target domain. In compliance with the guidelines of DCASE Task 5, we adopted the enhanced prototypical network with negative hard sampling as the new baseline for the DCASE 2024 Task 5 competition. Our goal is to stimulate future research on domain adaptation in the context of few-shot bioacoustic sound event detection.

\section{Acknowledgements} \label{sec:acknowledgement}
We would like to thank Haohe Liu for his valuable discussion on this work, and Hanna Pamula, Helen Whitehead and Frants Jensen for contributing the new data for the 2024 validation set. 
The research utilised Queen Mary's Apocrita HPC facility, supported by QMUL Research-IT, \url{http://doi.org/10.5281/zenodo.438045}. J. Liang is supported by the Engineering and Physical Sciences Research Council [grant number EP/T518086/1]. I. Nolasco is supported by the Engineering and Physical Sciences Research Council [grant number EP/R513106/1]. B. Ghani is supported by EU-funded projects MAMBO and GUARDEN. E. Benetos is supported by a RAEng/Leverhulme Trust Research Fellowship [grant number LTRF2223-19-106]. For the purpose of open access, the authors have applied a Creative Commons Attribution (CC BY) license to any Author Accepted Manuscript version arising.

\bibliographystyle{src/IEEEtran}
\bibliography{src/ref}

\begin{thebibliography}{10}
\providecommand{\url}[1]{#1}
\csname url@samestyle\endcsname
\providecommand{\newblock}{\relax}
\providecommand{\bibinfo}[2]{#2}
\providecommand{\BIBentrySTDinterwordspacing}{\spaceskip=0pt\relax}
\providecommand{\BIBentryALTinterwordstretchfactor}{4}
\providecommand{\BIBentryALTinterwordspacing}{\spaceskip=\fontdimen2\font plus
\BIBentryALTinterwordstretchfactor\fontdimen3\font minus \fontdimen4\font\relax}
\providecommand{\BIBforeignlanguage}[2]{{%
\expandafter\ifx\csname l@#1\endcsname\relax
\typeout{** WARNING: IEEEtran.bst: No hyphenation pattern has been}%
\typeout{** loaded for the language `#1'. Using the pattern for}%
\typeout{** the default language instead.}%
\else
\language=\csname l@#1\endcsname
\fi
#2}}
\providecommand{\BIBdecl}{\relax}
\BIBdecl

\bibitem{caiger2020decade}
P.~E. Caiger, M.~J. Dean, A.~I. DeAngelis, L.~T. Hatch, A.~N. Rice, J.~A. Stanley, C.~Tholke, D.~R. Zemeckis, and S.~M. Van~Parijs, ``A decade of monitoring atlantic cod gadus morhua spawning aggregations in massachusetts bay using passive acoustics,'' \emph{Marine Ecology Progress Series}, vol. 635, pp. 89--103, 2020.

\bibitem{gillings2021nocturnal}
S.~Gillings and C.~Scott, ``Nocturnal flight calling behaviour of thrushes in relation to artificial light at night,'' \emph{Ibis}, vol. 163, no.~4, pp. 1379--1393, 2021.

\bibitem{tuia2022perspectives}
D.~Tuia, B.~Kellenberger, S.~Beery, B.~R. Costelloe, S.~Zuffi, B.~Risse, A.~Mathis, M.~W. Mathis, F.~Van~Langevelde, T.~Burghardt \emph{et~al.}, ``Perspectives in machine learning for wildlife conservation,'' \emph{Nature communications}, vol.~13, no.~1, pp. 1--15, 2022.

\bibitem{ganchev2017computational}
T.~Ganchev, \emph{Computational bioacoustics: biodiversity monitoring and assessment}.\hskip 1em plus 0.5em minus 0.4em\relax Walter de Gruyter GmbH \& Co KG, 2017, vol.~4.

\bibitem{stowell2022computational}
D.~Stowell, ``Computational bioacoustics with deep learning: a review and roadmap,'' \emph{PeerJ}, vol.~10, p. e13152, 2022.

\bibitem{ghani2023global}
B.~Ghani, T.~Denton, S.~Kahl, and H.~Klinck, ``Global birdsong embeddings enable superior transfer learning for bioacoustic classification,'' \emph{Scientific Reports}, vol.~13, no.~1, p. 22876, 2023.

\bibitem{hamer2023birb}
J.~Hamer, E.~Triantafillou, B.~van Merrienboer, S.~Kahl, H.~Klinck, T.~Denton, and V.~Dumoulin, ``Birb: A generalization benchmark for information retrieval in bioacoustics,'' \emph{arXiv preprint arXiv:2312.07439}, 2023.

\bibitem{boudiaf2023search}
M.~Boudiaf, T.~Denton, B.~Van~Merri{\"e}nboer, V.~Dumoulin, and E.~Triantafillou, ``In search for a generalizable method for source free domain adaptation,'' in \emph{International Conference on Machine Learning}.\hskip 1em plus 0.5em minus 0.4em\relax PMLR, 2023, pp. 2914--2931.

\bibitem{nolasco2023learning}
I.~Nolasco, S.~Singh, V.~Morfi, V.~Lostanlen, A.~Strandburg-Peshkin, E.~Vida{\~n}a-Vila, L.~Gill, H.~Pamu{\l}a, H.~Whitehead, I.~Kiskin \emph{et~al.}, ``Learning to detect an animal sound from five examples,'' \emph{Ecological informatics}, vol.~77, p. 102258, 2023.

\bibitem{Li2022-3}
R.~Li, J.~Liang, and H.~Phan, ``Few-shot bioacoustic event detection: Enhanced classifiers for prototypical networks,'' in \emph{Proceedings of the 7th Detection and Classification of Acoustic Scenes and Events 2022 Workshop (DCASE2022)}, Nancy, France, November 2022.

\bibitem{Liang2022}
J.~Liang, H.~Phan, and E.~Benetos, ``Leveraging label hierachies for few-shot everyday sound recognition,'' in \emph{Proceedings of the 7th Detection and Classification of Acoustic Scenes and Events 2022 Workshop (DCASE2022)}, Nancy, France, November 2022.

\bibitem{liang23c_interspeech}
J.~Liang, X.~Liu, H.~Liu, H.~Phan, E.~Benetos, M.~D. Plumbley, and W.~Wang, ``{Adapting Language-Audio Models as Few-Shot Audio Learners},'' in \emph{Proc. INTERSPEECH 2023}, 2023, pp. 276--280.

\bibitem{morfi2021few}
V.~Morfi, I.~Nolasco, V.~Lostanlen, S.~Singh, A.~Strandburg-Peshkin, D.~Benvent, and D.~Stowell, ``Few-shot bioacoustic event detection: A new task at the dcase 2021 challenge.''

\bibitem{nolasco2022few}
I.~Nolasco, S.~Singh, E.~Vidana-Villa, E.~Grout, J.~Morford, M.~Emmerson, F.~Jensens, H.~Whitehead, A.~Strandburg-Peshkin \emph{et~al.}, ``Few-shot bioacoustic event detection at the dcase 2022 challenge,'' \emph{arXiv preprint arXiv:2207.07911}, 2022.

\bibitem{nolasco2023few}
I.~Nolasco, B.~Ghani, S.~Singh, E.~Vida{\~n}a-Vila, H.~Whitehead, E.~Grout, M.~Emmerson, F.~Jensen, I.~Kiskin, J.~Morford \emph{et~al.}, ``Few-shot bioacoustic event detection at the dcase 2023 challenge,'' \emph{arXiv preprint arXiv:2306.09223}, 2023.

\bibitem{Liu2022a}
H.~Liu, X.~Liu, X.~Mei, Q.~Kong, W.~Wang, and M.~D. Plumbley, ``Surrey system for dcase 2022 task 5 : Few-shot bioacoustic event detection with segment-level metric learning,'' DCASE2022 Challenge Technical Report, Tech. Rep., 2022.

\bibitem{robinson2020contrastive}
J.~Robinson, C.-Y. Chuang, S.~Sra, and S.~Jegelka, ``Contrastive learning with hard negative samples,'' \emph{arXiv preprint arXiv:2010.04592}, 2020.

\bibitem{arnold2007comparative}
A.~Arnold, R.~Nallapati, and W.~W. Cohen, ``A comparative study of methods for transductive transfer learning,'' in \emph{Seventh IEEE international conference on data mining workshops (ICDMW 2007)}.\hskip 1em plus 0.5em minus 0.4em\relax IEEE, 2007, pp. 77--82.

\bibitem{snell2017prototypical}
J.~Snell, K.~Swersky, and R.~S. Zemel, ``Prototypical networks for few-shot learning,'' 2017.

\bibitem{Liang2024}
J.~Liang, H.~Phan, and E.~Benetos, ``Learning from taxonomy: Multi-label few-shot classification for everyday sound recognition,'' in \emph{ICASSP 2024 - 2024 IEEE International Conference on Acoustics, Speech and Signal Processing (ICASSP)}, 2024, pp. 1--5.

\end{thebibliography}

\end{document}